\documentclass[prl]{revtex4}%
\pdfoutput=1

\usepackage{epsfig}
\usepackage{graphics}

\begin{document}

\title{Quantum Indeterminacy of Emergent Spacetime}

\author{Craig J. Hogan}
\affiliation{University of Washington,  Seattle and Max-Planck-Institut f\"ur Astrophysik, Garching bei M\"unchen}

\begin{abstract}
It is shown that nearly-flat  3+1D spacetime emerging from  a dual quantum field theory in 2+1D   displays quantum fluctuations from classical Euclidean geometry on macroscopic scales.  A covariant holographic mapping is assumed,  where plane wave states with wavevector $\vec k$  on a  2D surface map onto classical null trajectories  in the emergent third dimension at an angle $\vec\theta=l_P\vec k$ relative to the surface element normal, where $l_P$ denotes the Planck length. Null trajectories in the 3+1D world then display   quantum uncertainty of angular orientation,  with standard deviation $\Delta\theta=\sqrt{l_P/z}$ for longitudinal propagation distance $z$ in a given frame.   The quantum complementarity of transverse position at macroscopically separated events along null trajectories corresponds to  a geometry that is not completely classical, but  displays observable  holographic quantum noise. A statistical estimator of the fluctuations  from Euclidean behavior is given for a simple thought experiment based on measured sides of   triangles. The effect can be viewed as sampling noise due to the limited degrees of freedom of such a theory,  consistent with   covariant bounds on entropy. 
\end{abstract}
\pacs{04.60.-m}
\maketitle
\section{Introduction}
It is not understood  how a fundamentally quantum world creates the appearance, to internally constituted observers,  of an approximately classical 3+1D classical spacetime, within which quantum fields operate.  There are however indications that general relativity  is in some sense ``holographic'' and that spacetime somehow emerges from a quantum theory with lower dimensionality.  Examples exist where a conformal field theory with  D spatial dimensions appears to be dual to a (supersymmetric) quantum field theory with gravity in D+1 spatial dimensions\cite{Maldacena:1997re,Witten:1998zw,Aharony:1999ti,Alishahiha:2005dj,Horowitz:2006ct}.
Arguments based on black  hole thermodynamics and evaporation\cite{Bekenstein:1972tm,Bardeen:gs,Bekenstein:1973ur,Bekenstein:1974ax,Hawking:1975sw,Hawking:1976ra,'tHooft:1985re,Susskind:1993if},   string theory\cite{Strominger:1996sh},  thermodynamics  in nearly-flat space\cite{Jacobson:1995ab},  and classical relativity\cite{Padmanabhan:2006fn,Padmanabhan:2007en,Padmanabhan:2007xy,
Padmanabhan:2007tm}, have led many authors to suspect that a region of  3+1D spacetime with gravity  and the fields within it may have a  dual description in terms of null surfaces swept out by a 2+1D quantum    theory with a cutoff at the Planck length $l_P$ \cite{Padmanabhan:2006fn,Padmanabhan:2007en,Padmanabhan:2007xy,
Padmanabhan:2007tm,'tHooft:1993gx,Susskind:1994vu,'tHooft:1999bw,Bigatti:1999dp,Bousso:2002ju}.  This paper shows that such holographic theories generally display directly measurable quantum departures from classical general relativity.

Recently, the author has argued\cite{Hogan:2007rz,Hogan:2007hc,Hogan:2007ci} that  a nearly-flat  holographic  spacetime   approaches classical behavior  on macroscopic scales   in a distinctive way that  displays  small but detectable quantum fluctuations from  Euclidean geometry even at macroscopic separations. Specifically, the   classical metric at widely separated points along any null trajectory displays a quantum complementarity of transverse position.  Conversely, at small separations, angles become increasingly poorly defined  until at the Planck scale the third dimension melts into the behavior of a 2+1D theory.  This paper shows how this behavior arises simply from the  geometry of the holographic mapping of the 2+1D  states into 3+1D. This exercise helps to define the class of theories that have general relativity as an approximate classical limit, and that display the phenomena of  holographic indeterminacy and noise.

\section{holographic  mapping of null surfaces}

\begin{figure}
\epsfysize=4.5in 
\epsfbox{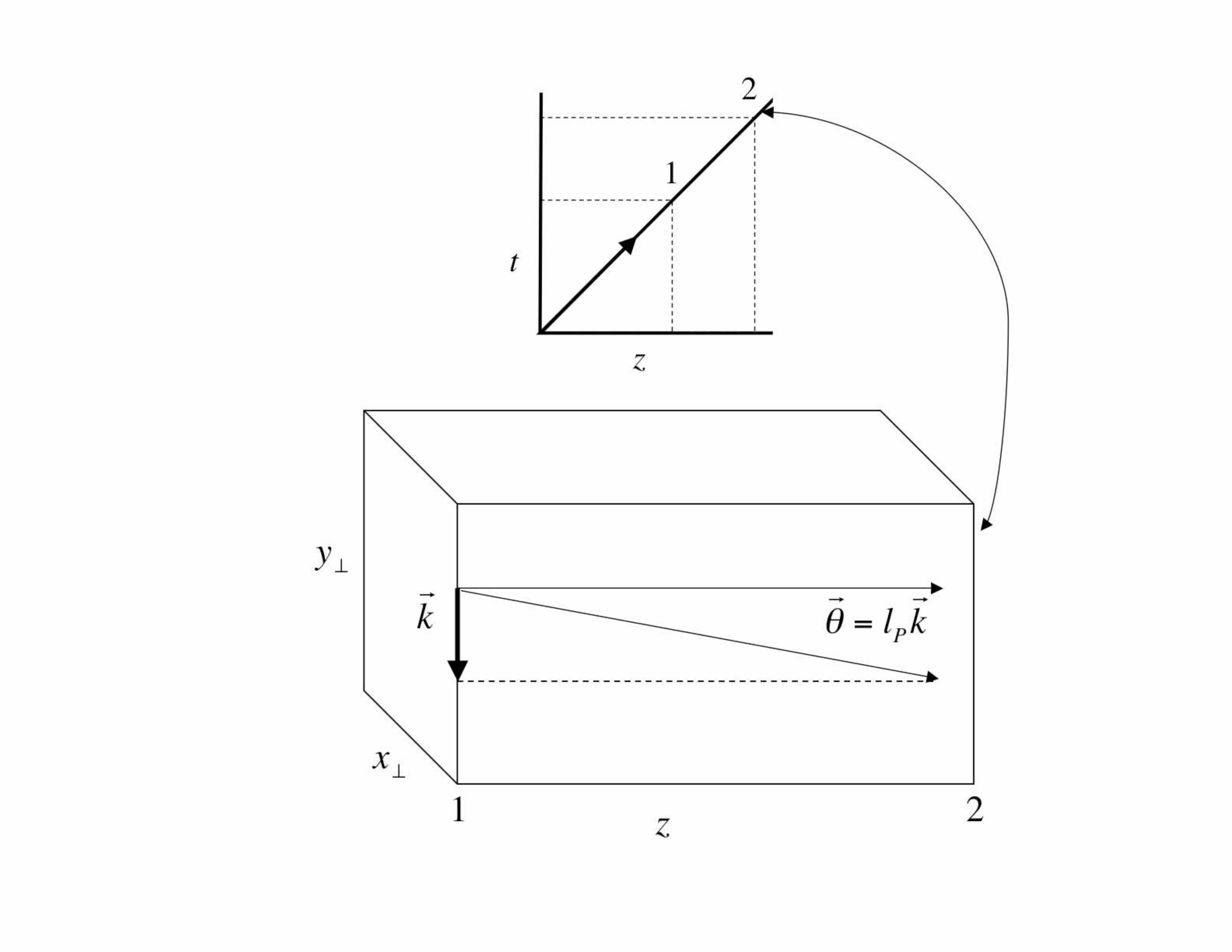} 
\caption{ \label{projections}
The behavior of   a quantized, local theory on a 2D   surface is interpreted as a 3+1D spacetime.  Above, a spacetime diagram in  a particular frame in   the emergent 3+1D spacetime; each point is a 2D surface element in the $x_\perp,y_\perp$ plane.     In the dual holographic interpretation of the fields' behavior, a  surface element   sweeps out  a  null 3D volume (along the null ray shown).  Below, the 3D spatial volume swept out by the surface is shown projected onto the spacelike $z$ axis in a particular frame.  Plane wave states with wavevector $\vec k$  in 2D map onto classical null trajectories  in the emergent third dimension at an angle $\vec\theta=l_P\vec k$ relative to the surface element normal. A classical path  localized in the transverse direction corresponds to a state with finite support on the  2D surface;  as a result, null paths in the emergent 3+1D world display angular indeterminacy.}
\end{figure}

Suppose that the basic element  is a quantized field theory on a two-dimensional  spacelike surface,  with a Planck wavelength UV cutoff.   The surface can be propagated on a timelike world-line, in which case the theory describes the time evolution of a 2+1D quantum world. If the theory has a dual interpretation with an extra spatial dimension,     states in the  2+1D world also  describe states of a null surface  in a 3+1D spacetime with an extra emergent third spatial dimension (see Figure \ref{projections}).  In the 3+1D world, each 2D surface element sweeps  through the new ``virtual'' spatial dimension at the speed of light, in a direction normal to the surface.

Now suppose that the quantum states of the  2D theory map onto the third  spatial dimension in a particular way:  information about the third dimension is encoded in   the 2D fields in a way similar to an actual hologram, which forms an interferogram of 3D wavefronts  on a surface.  This kind of  2+1D theory   describes a 3+1D world   with a  particular kind of  information about transverse correlations and positions at large separations.  Specifically the discussion below assumes that  {\em  plane wave states with wavenumber $\vec k$  in 2D map onto classical null trajectories  in the emergent third dimension at an angle $\vec\theta=l_P\vec k$ relative to the surface element normal.}  Here the two components  of  $\vec\theta$ correspond to the projected $\vec x= (x_\perp,y_\perp)$ plane.  This choice of mapping, besides being motivated by  the holographic metaphor, seems uniquely dictated by the requirements of linearity, covariance,  and independence of axes. 
The choice of momentum normalization  is fixed by requiring that propagation along an  emergent (null) trajectory   occurs via Planck scale operators; that is, a family of null trajectories parallel to the normal $z$ axis    corresponds to     a  plane wavefunction with  wavelength $l_P$.  This choice is dictated by the need for $l_P$ to be the only fundamental length unit; in any 3+1D frame, the theory must only depend  on pure numbers, hence on distances in units of $l_P$.
Similarly, the Planck length sets the scale of momenta in the plane of the surface element, transverse to the direction of motion.

For an infinite 2D surface, an angular orientation in 3D maps onto a point in 2D momentum space, where a small 3D angle relative to normal corresponds to  a long wave in 2D. Such  a state encodes no depth information, only an orientation. On the other hand,  the orientation specified by a finite region     of 2D surface  is imperfectly defined, since it is limited by diffraction.  In  an actual hologram,     longitudinal displacement in the $z$ dimension shows up in a stereo image: different apertures in the plane of the hologram encode different directions to the same object, albeit imperfectly. Similarly, information about the third, emergent dimension at longitudinal separation $z$ in a given frame is encoded by  the holographic pattern of fields in 2D at the characteristic diffraction scale, $\sqrt{zl_P}$; an aperture of this size has a diffraction spot at distance $z$ about equal to its own size. Depth information is encoded in 2D field correlations on this scale and longer.

While an angular orientation for a plane wave can be precisely defined by fields over an infinitely large 2D patch, there is no corresponding mapping for a metric defined  by 3+1D particle paths.   Localization in the transverse directions necessarily creates indeterminacy in  the orientation of any  classical trajectory.  The emergence of a metric defined by classically observable positions  requires specification of paths in 3+1D which are subject to     indeterminacy on the diffraction scale $\sqrt{zl_P}$.  

\section{quantum theory of holographic indeterminacy} 

Holographic indeterminacy can be described as a result of quantum particle/wave complementarity. A classical metric  defines distances between points on a manifold and therefore corresponds to  definite particle trajectories specified by position observables. 
In  3+1D space, a  classical null trajectory is specified by  an orientation, the $z$ axis normal to a surface element,  and a transverse position $(x_\perp,y_\perp)$.  The latter observable corresponds to  a particle wavefunction with support over a limited patch of surface, which creates a Heisenberg uncertainty in transverse momentum, and therefore in distant transverse position relative to the $z$ axis. Thus position observables at events widely separated in $z$ become conjugate  quantum operators with  associated indeterminacy.

 To describe this effect more precisely, we switch to the complementary particle description of a classical null trajectory \cite{Hogan:2007rz,Hogan:2007hc,Hogan:2007ci}.  The ``particle'' in this description does not refer as usual to a physical particle in a fixed metric.
The particle description instead corresponds to states describing a definite metric and definite classical paths,  which are complementary (in the quantum sense) to the wave description.     On surface $1$  a  particle  obeys the   Heisenberg commutation relation between conjugate momentum and position observable operators along the transverse $x$-axis:
\begin{equation}\label{heisenberg}
[\hat x_\perp(z_1),\hat p_\perp(z_1)]=-i\hbar.
\end{equation}
The transverse momentum  $p_{x\perp}(z_1)$ of a Planck momentum  particle on surface 1 is related to its transverse position displacement on surface 2 by the angular deflection,
\begin{equation}\label{deflection}
 p_{x\perp}(z_1)l_P/\hbar= x_\perp(z_2)/ (z_2-z_1).
\end{equation}
Combining equations (\ref{heisenberg}) and (\ref{deflection}) 
yields a commutation relation between transverse position operators, 
\begin{equation}\label{commute}
[\hat x_\perp(z_1),\hat x_\perp(z_2)]=-i l_P (z_2-z_1),
\end{equation}
where Planck's constant $\hbar$ has dropped out.
This formula specifies the complementarity of the transverse position observables  along a null trajectory,   and thereby expresses the uncertainty of the emergent null trajectory itself.
In the usual way, the indeterminacy (Eq. \ref{commute}) yields a Heisenberg uncertainty relation,
\begin{equation}\label{uncertainty}
\Delta x_\perp(z_1)\Delta x_\perp(z_2)>l_P (z_2-z_1)/2,
\end{equation}
where $ \Delta x_\perp(z_1),\Delta x_\perp(z_2)$ denote the standard deviations on surfaces 1 and 2 of the distributions of position measurements, describing  departures from a classical null ray.  The standard deviation $\Delta x_\perp$ of the  difference in relative transverse positions   is then given by $\Delta x_\perp^2=  \Delta x^2_\perp(z_1)+\Delta x^2_\perp(z_2)$; it has a minimum value when $ \Delta x_\perp(z_1)=\Delta x_\perp(z_2)$,
so there is a ``holographic uncertainty principle'' for relative  transverse positions   at events of null spacetime separation and spatial separation $z$ in a given frame:
\begin{equation}\label{uncertainty2}
\Delta x_\perp^2>l_Pz.
\end{equation}
From this we also derive a minimum uncertainty in angular orientation of any null ray of length $L$ along  each transverse axis:
\begin{equation}\label{angledelta}
\Delta \theta_x> \sqrt{l_P/L},\ \  \Delta \theta_y>  \sqrt{l_P/L}.
\end{equation}

As expected for an emergent spacetime,  angular uncertainty   increases with smaller $L$, so that     a classical spatial direction is   ill defined at the Planck scale and only becomes well defined after many Planck lengths of propagation.  
Angles become better defined at larger macroscopic separation--- in this sense the world becomes ``more classical'' as it becomes on larger scales  ``more three dimensional.''  What is surprising and new is that  transverse positions in absolute terms actually  become less well defined at larger separations.   Transverse positions of  macroscopically separated bodies    at some level     do not exist as separately observable quantities, but are complementary: knowledge of one precludes accurate knowledge of the other.   Even macroscopic spacetime exhibits quantum departures from classical Euclidean behavior.  In this sense the emergent spacetime radically differs from the usual classical limit.

\section{experimental predictions}
The quantum fluctuations about  Euclidean geometry can be observed from inside the 3+1D spacetime. Suppose light beams are emitted from an event in two different spatial directions. The length and  transverse distance between the beams can be measured (by interferometry, say). In a Euclidean space this measurement determines the angle $\theta_1$ between them and all subsequent measurements yield the same result. However,  in emergent spacetime, subsequent measurements of the same beams yield different angles.  Indeed there is no ``true'' classical value of $\theta_1$ for finite beams.

\begin{figure}
\epsfysize=4in 
\epsfbox{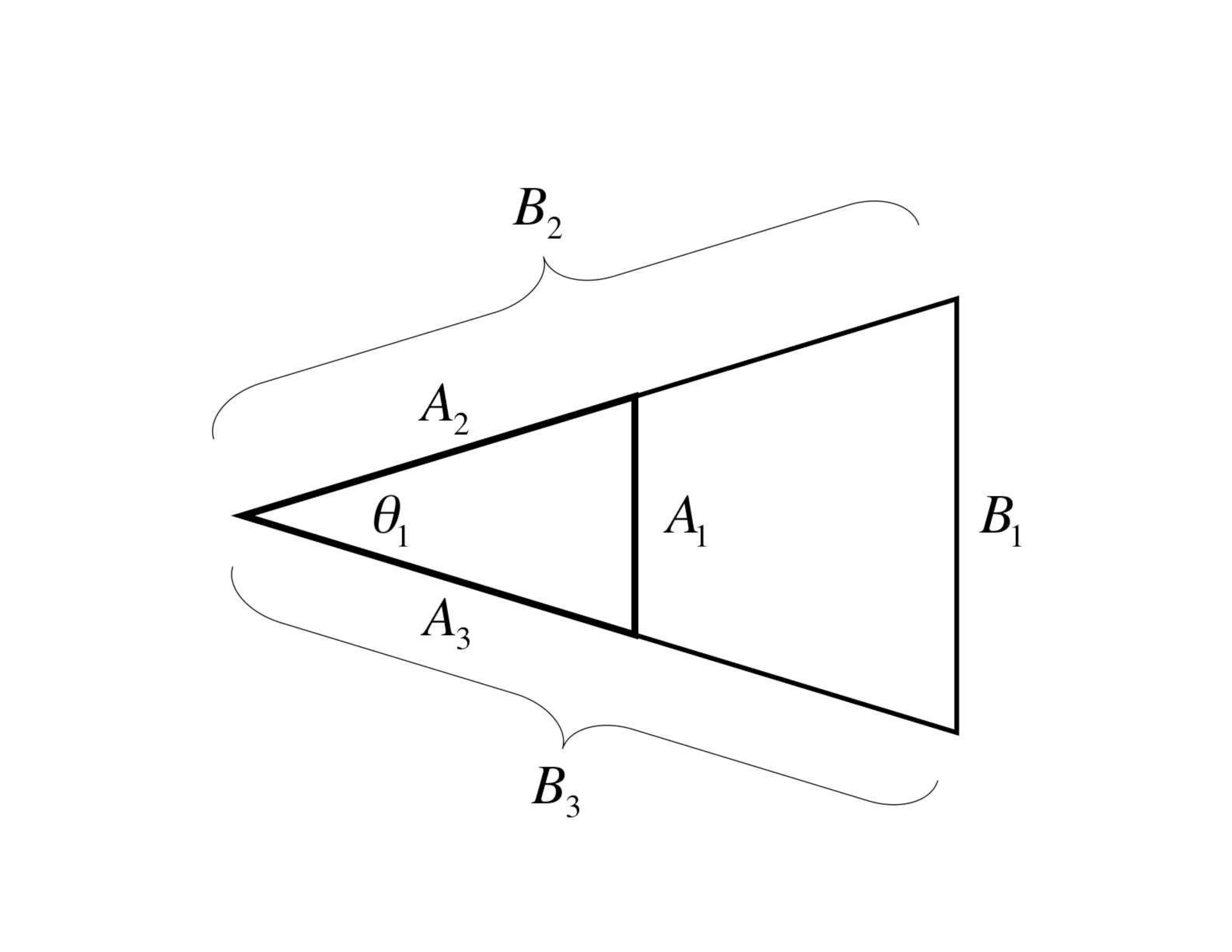} 
\caption{ \label{triangles}
Triangles sharing a common angle $\theta_1$ for two null rays.  Estimates of  $\theta_1$ based on measurements of the sides at different times exhibit holographic noise.}
\end{figure} 

It is useful to illustrate  the effect quantitatively with a concrete thought experiment (see Fig. \ref{triangles}).  Let two rays be sent out from an event, separated by an angle $\theta_1$.  At some later time the two rays are intercepted at two other events, defining a triangle, $A$.  In Euclidean space, the angle $\theta_1$ and the length of three sides of triangle $A_1,A_2,A_3$ are related by 
\begin{equation}
\cos^2\theta_1= {{A_2^2+A_3^2-A_1^2}\over {2 A_2A_3}}.
\end{equation}
For two such triangles $A$ and $B$,  measured via the same rays at different events and  sharing the common angle $\theta_1$, we can define a measurable quantity based on radial measurements of the triangle sides,
\begin{equation}
Q_{AB}^2={{A_2^2+A_3^2-A_1^2}\over {2 A_2A_3}}
-{{B_2^2+B_3^2-B_1^2}\over {2 B_2B_3}},
\end{equation}
that has the property of vanishing in Euclidean space.  However, in the presence of holographic indeterminacy the quantum departure from classical null rays leads to  nonzero fluctuations in $Q_{AB}^2$, because   $A_1$ and $B_1$ are complementary observables and cannot both be measured precisely. 
The   variance   in the distribution--- a measure of holographic noise--- is
given by
\begin{equation}
\Delta Q_{AB}^2= {\Delta A_1^2\over 2 A_2A_3} +{\Delta B_1^2\over 2 B_2B_3},
\end{equation}
where $\Delta A_1$ and $\Delta B_1$ are the 
projected uncertainties  of transverse distance (from Eq. \ref{uncertainty2}) on the measured far legs of the triangle. For a nearly isoceles triangle and fractionally small differences in the sizes of triangles $A$ and $B$,   the sum of these is  $\cos (\theta_1/2)\sin(\theta_1/2)$ times the transverse relative uncertainty $\Delta x_\perp$ arising from the longitudinal distance between vertices on each ray, $B_2-A_2$ and $B_3-A_3$, where the transverse positions are measured: 
 \begin{equation}\label{trianglenoise}
\Delta Q_{AB}^2\simeq {{ \cos^2(\theta_1/2)\sin^2(\theta_1/2) [(B_2-A_2)+(B_3-A_3)] l_P}\over 2 A_2 A_3}.
\end{equation}
The variance represents a decoherence  that accumulates linearly in propagation distance along the rays between the observations at the events that define the triangles,  resembling a random walk with transverse Planck length steps for each Planck length of propagation.
  At the same time the scaling is such that the dimensionless  variance   decreases as the size of the triangles increases, as expected for a system that   gradually becomes more classical as it gets larger. Equation (\ref{trianglenoise}) represents an observational prediction for a statistical  distribution of measurements that should vanish in the absence of holographic noise.   Similar predictions can be defined for more realistic  experiments that precisely measure relative  transverse positions using interferometry.

Another way of summarizing this effect is that the uncertainty leads to detectable quantum position noise with power spectral density $\approx l_P$, corresponding to holographic  quantum shear noise in the metric. The flat spectrum  of dimensionless metric perturbations,
 \begin{equation}
h_{H} \simeq  \sqrt{l_P/c} = {2.3 \times 10^{-22}{\rm Hz}^{-1/2}},
\end{equation}
 implies holographic noise, or quantum fluctuations about a Euclidean geometry,   on all scales \cite{Hogan:2007rz,Hogan:2007hc,Hogan:2007ci}. Interferometry with current technology appears capable of detecting noise at this level over a wide range of scales.

\section{Information limits and macroscopic quantum weirdness}

The number of degrees of freedom does not depend on whether the theory is intepreted as 2+1D or 3+1D. 
The uncertainty in   orientations or angles translates into a limit on the number of distinguishable angular orientations of plane-wave modes of   quantum fields in 3+1D.  
In the 3D+1 world, taking both transverse directions into account, and counting states by assuming Nyquist sampling on a sphere (that is, states or degrees of freedom separated by two standard deviations in each transverse direction $x$ and $y$), the number of distinguishable orientations for rays   of length $L$  in a sphere of  radius $L$ is 
$ N_\theta(L)<{{4\pi}/{4\Delta \theta_x \Delta \theta_y}}=\pi L/l_P.$  This corresponds to    the number of distinguishable wavevector directions for field modes confined  to the volume. 
The number of quantum degrees of freedom up to the Planck scale is  $\approx L/l_P$ field modes per direction, so the number of degrees of freedom is $
\approx N_\theta(L)L/l_P<\pi (L/l_P)^2$--- as expected, one quarter of the projected area of of the sphere in Planck units.
This    estimate agrees (up to a numerical factor of the order of unity, depending on the exact nature of the Planck cutoff) with the maximum number of degrees of freedom allowed by covariant entropy bounds\cite{Bousso:2002ju}. Holographic noise can be thought of as ``sampling noise'' due to the limited number of degrees of freedom.  The number of distinguishable wavevector directions in the same volume is  much smaller (by a factor $\simeq L/l_P$) than the number in standard field theory quantized in 3D with a Planck scale cutoff,   and the macroscopic quantum fluctuations in the metric are correspondingly larger.

Because the emergent 3+1D spacetime is defined by null paths,  it preserves statistical Lorentz invariance: a 2+1D dual interpretation can be found with respect to any choice of  reference frame. 
Of course, the full symmetry (as well as the unitary evolution) refers to the full quantum state:   the specific observed classical metric into  which the state ``collapses'' due to a position measurement depends on the  frame and the spatial location in which the position measurement  is made.  Measurement of a position anywhere in spacetime has the effect of selecting a  branch of  the spacetime wavefunction that is an eigenstate of that position.  Since local transverse position observables (almost) commute, this branch behaves near the measurement event almost like a classical metric, but it  is actually a superposition of states for quantities--- such as distant, transverse positions on the future light cone of the measurement--- that do not commute with that observable. Although this ``quantum weirdness''
of macroscopic spacetime is small and would not have been detected up to now, it is probably large enough to be within reach of realistic experiments able to measure small relative transverse positions at macroscopic separations \cite{Hogan:2007rz,Hogan:2007hc,Hogan:2007ci}.

\acknowledgements
The author is grateful for support from the Alexander von Humboldt Foundation.

\end{document}